\documentstyle[sprocl]{article}
\newcommand{\beq}{\begin{equation}}
\newcommand{\eeq}{\end{equation}}
\newcommand{\ald}{\dot \alpha}
\newcommand{\bed}{\dot \beta}

\newcommand{\ej}{\cal E}
\newcommand{\hej}{\hat {\ej}}

\def\Journal#1#2#3#4{{#1} {\bf #2}, #3 (#4)}

\def\PLB{{\em Phys. Lett.}  B}

\def\be{\begin{equation}}
\def\ee{\end{equation}}
\def\bea{\begin{eqnarray}}
\def\eea{\end{eqnarray}}
\begin{document}

\title{CLASSICAL SOLUTIONS GENERATING TREE FORM-FACTORS IN
YANG-MILLS, SIN(H)-GORDON AND GRAVITY}

\author{K. G. SELIVANOV}

\address{ITEP, Cheremushkinskaya 25, 117218 Moscow,\\ 
Russia\\E-mail: selivano@heron.itep.ru}


\maketitle\abstracts{Classical solutions generating tree form-factors
are defined and constructed in various models}

\section{Definition and Motivation}
In this my talk I am going to describe classical solutions
which are generating functions for tree form-factors in 
the corresponding quantum theory. My talk will be based on
works done in collaboration with A.Rosly ~\cite{KS1} $^{-}$ \cite{KS5}.
I should also cite related papers by W.Bardeen ~\cite{Ba} and
by V.Korepin and T.Oota ~\cite{KO}. These classical solutions have 
been called {\it perturbiners} and have been constructed in various
models. I believe that in this audience the most convenient is
to start just with definition.\\
\underline{Definition}\\
Consider a nonlinear field field equation:
\begin{equation}
\label{1}
({\partial}^{2}+m^{2}){\phi}+\lambda{\phi}^{2}+\ldots=0
\end{equation}
(I take here the case of scalar field,
generalizations are trivial).\\
Take a solution of the corresponding free field equation,
$({\partial}^{2}+m^{2}){\phi}=0$,
in the form of linear combination of a set of plane waves,
${\phi}^{(1)}=\sum_{j=1}^{N} a_{j}e^{ik_{j}x}=
\sum_{j} {\ej}_{j}$
(in the non-scalar case there is a polarization factor etc.
in front of the plain waves).\\
Assume:\\
a)non-resonantness:
$\left(\sum_{j}n_{j}k_{j} \right)^{2} \neq m^{2}$, 
when the sum contains more than one term ($n_{j}=0,1$);\\
b)nilpotency:
$a_{j}^{2}=0$.\\
{\it Perturbiner} is a complex solution of Eq.~(\ref{1})
of the following type:
\begin{eqnarray}
\label{6}
{\phi}_{ptb}(x,\{k_{j}\},\{a_{j}\})={\phi}^{(1)}(x,\{k_{j}\},\{a_{j}\})+
{\rm higher}\nonumber\\
{\rm order\; terms\;in\;the\;plane \;waves}\; \{{\ej}_{j}\}\;
{\rm entering}\;{\phi}^{(1)}
\end{eqnarray}
Solution of this type obviously exists and is unique.
Due to the nilpotency condition there is a finite
number of terms in Eq.~(\ref{6}) and every term is well defined because
the operator $({\partial}^{2}+m^{2})$ from Eq.~(\ref{1}) is invertible
in the space of polynomials in $\{{\ej}_{j}\},\;j=1,\ldots,N$ in the 
non-resonantness assumption. In gauge theories the uniqueness
takes place after gauge fixing or, equivalently, modulo gauge 
transformations.\\
\underline{Motivation}\\
The perturbiner is a generating function for the so-called form-factors
in the tree approximation,\\
${\phi}^{ptb}(x, \{k\}, \{a\})=\sum_{d=1}^{N}\sum_{j_{1} \ldots j_{d}}
<k_{j_{1}}, \ldots , k_{j_{d}}|{\phi}(x)|0>_{tree}
a_{j_{1}} \ldots a_{j_{d}}$.\\
The nilpotency is equivalent to excluding the form-factors with
identical particles (no loss of generality provided the perturbiner
is known for any $N$).\\
The non-resonantness warrants that there are no internal lines
on-shell.\\
\underline{Notice}:\\
Classical text-books on QFT, e.g. books~\cite{FS}$^{,}$\cite{IZ}, 
contain chapters about
classical solutions generating tree amplitudes but they use 
different definitions (asymptotic Feynman-type boundary conditions)
and do not give any explicit examples.

In the case of our definition the perturbiner is constructed explicitely
in all cases when the field equations admit a zero-curvature representation
with one-dimensional auxiliary (spectral) space. Actually, in terms of the
zero-curvature representation the construction is essentially the same 
even though the original models look very different ( 4d Yang-Mills and
2d sin(h)-Gordon).

\section{Yang-Mills}
Yang-Mills equations do not have one-dimensional zero-curvature 
representation. Therefore we consider not generic Yang-Mills perturbiner,
but the one which generates only positive helicity form-factors,
$<^{+}k_{1}, \ldots, ^{+}k_{N}|A_{\mu}(x)|0>$, that is the solution
of Yang-Mills equations of the following type,\\
$A_{\mu}^{ptb}(x,\{k_{j}\},\{a_{j}\})=\sum_{j=1}^{N} \epsilon^{j}_{\mu}t^{j}
a_{j}e^{ik_{j}x}+
{\rm higher\;order\;terms\;in\;the\;plane \;waves}\\ 
\{{\ej}_{j}=a_{j}e^{ik_{j}x}, j=1,\ldots,N\}$,
where $\epsilon^{j}_{\mu}$ are positive helicity polarizations,
$t^{j}$ are color matrixes.
{\bf Such $A_{\mu}^{ptb}$ obeys self-duality equations.}
Indeed, linearized self-duality equation is obviously equivalent to the
positive helicity condition, since both assume that ``electric field''
is equal to $i \cdot$ ``magnetic field''. Beyond the linear approximation, 
solutions of the self-duality equations are solutions of the Yang-Mills
equations as well, and for the Yang-Mills equations the solution of the type of
perturbiner is unique (modulo gauge transformations, see above).
{The self-duality equations do have a zero-curvature representation
with one-dimensional auxiliary space - the twistor representation.}\\
Notice: we get integrability not by substituting the theory, we just
reduce the set of magnitudes we pretend to compute.\\
\underline{Solving self-duality equations:}~\cite{Ward}\\
It is convenient here to use the spinor notations, so that all
vector objects have two spinor indices, e.g. the partial derivative
$\partial_{\alpha \ald}=\frac{\partial}{\partial x^{\alpha \ald}}$,
the connection-form $A_{\alpha \ald}$, and the connection itself,
$\nabla _{\alpha \ald}=\partial_{\alpha \ald}+A_{\alpha \ald}$,
where $\alpha, \ald =1,2$.\\
The curvature form, 
\beq
\label{9}
F_{\alpha \ald \beta \bed}=[\nabla_{\alpha \ald},\nabla_{\beta \bed}]
\eeq
has four spinor indices, and being antisymmetric with respect to the 
transposition of pairs of indices, decomposes as follows:
\beq
\label{10}
F_{\alpha \ald \beta \bed}=\varepsilon_{\alpha \beta}F_{\ald \bed}+
\varepsilon_{\ald \bed} F_{\alpha \beta}
\eeq
where $\varepsilon$'s are the standard antisymmetric tensors and 
$ F_{\alpha \beta},F_{\ald \bed}$ are some symmetric tensors.
The first term in r.h.s. of Eq.~(\ref{10}) can be identified as
self-dual part of the curvature and the second term - as antiself-dual one.\\
Introduce now a couple of complex numbers, ${\rho}^{\alpha}, \alpha=1,2$,
which can be viewed on as homogeneous coordinates on auxiliary $CP^{1}$
space. Contracting couple of $\rho$'s with un-dotted indices in Eq.~(\ref{10})
one sees that the condition
$\rho^{\alpha}\rho^{\beta}F_{\alpha \ald \beta \bed}=0\;
{\rm identically\;in}\;\rho$,
is equivalent to the statement that $F_{\alpha \ald \beta \bed}$
is self-dual. On the other side, contracting couple of $\rho$'s with 
un-dotted indices in 
Eq.~(\ref{9}), one obtains the zero-curvature
representation of the self-duality,
$[\nabla_{ \ald},\nabla_{\bed}]=0\; {\rm at\, any}\; \rho^{\alpha},\alpha=1,2$
where $\nabla_{ \ald}=\rho^{\alpha}\nabla_{\alpha \ald}$. Thus, if one 
introduces
\begin{equation}
\label{13}
A_{\ald}=\rho^{\alpha}A_{\alpha \ald},\; 
\partial_{\ald}=\rho^{\alpha}\partial_{\alpha \ald}
\end{equation}
any self-dual connection form can be (locally) represented as
\begin{equation}
  \label{14}
A_{\ald}=g^{-1} \partial_{\ald} g  
\end{equation}
where $g$ is a group valued function of $\rho$ and $x$. All the non-triviality 
of the self-duality equation is now encoded in the condition that $g$ must 
depend on ${\rho}$ in such a way that $A_{\ald}$ is a polynomial of degree 1
in ${\rho}$, as in Eq.(\ref{13}). If $g$ is $\rho$-independent, it is a pure 
gauge transformation, as it is seen from  Eq.(\ref{14}).

The above condition on $\rho$-dependence of $g$ is equivalent to condition 
that $g$ is a homogeneous meromorphic function of $\rho$ of degree 0 such
that $A_{\ald}$ from Eq.(\ref{14}) is a homogeneous holomorphic function
of $\rho$ of degree 1 (a homogeneous holomorphic function
of $\rho$ of degree 1 is necessary just linear in $\rho$, as in 
Eq(\ref{13}).). Notice, that
nontrivial (not a pure gauge) $g$ necessary has singularities in $\rho$,
since if it is regular homogeneous of degree 0, then it is just 
$\rho$-independent, that is , a pure gauge.\\
\underline{One-particle solution:}\\
Consider the case when there is only one plane wave,
$A_{\alpha \ald}= \epsilon_{\alpha \ald}tae^{ik_{\beta \bed}x^{\beta \bed}}$.
The momentum of the particle, $k_{\beta \bed}$, is a light-like
four-vector, therefore it decomposes into a product of two spinors:
\beq
\label{16}
k_{\alpha \ald}=\ae_{\alpha} \bar{\ae}_{\ald}
\eeq
Due to the nilpotency of the parameter $a$ all equations are automatically 
linearized in the one-particle case.
The linearized self-duality equation,\\
$\varepsilon^{\ald \bed}\epsilon_{(\alpha \ald} k_{\beta) \bed}=0$,
assumes that the polarization $\epsilon_{(\alpha \ald}$ is also
light-like four-vector and its decomposition into a product
of two spinors contains the same dotted spinor as $k_{\beta \bed}$
in Eq.~(\ref{16}),
$\epsilon_{\alpha \ald}=q_{\alpha} \bar{\ae}_{\ald}$.
$q_{\alpha}$ is a reference spinor defined up to the 
so-called on-shell gauge freedom, $q_{\alpha}\sim q_{\alpha}+C\ae_{\alpha}$.
From the linearized version of Eq.~(\ref{14}),
$A_{\ald}=\partial_{\ald} g$,
one easily finds that 
\beq
\label{20}
g=1+\frac{(\rho q)}{(\rho \ae)} \hej
\eeq
where ${\ej}=a e^{i k_{\beta \bed} x^{\beta \bed}},  {\hej}=t {\ej}$ and
the brackets with two spinors, like $(\rho \ae)$, here and below stand
for contraction of the spinors with the $\varepsilon$-tensor,
$(\rho \ae)=\varepsilon^{\alpha \beta} \rho_{\alpha} \ae_{\beta}$
(indices of the spinors are raised and lowered with the 
$\varepsilon$-tensors).\\
Notice the simple pole of $g$ in Eq.~(\ref{20}) at 
$\rho_{\alpha}=\ae_{\alpha}$ which is absent in $A_{\ald}$.\\
\underline{$N$-particle solution:}\\
So our problem is to find $g^{ptb}(\rho,\{{\ej}_{j}\},\{k_{j}\})$ - polynomial
in $\{{\ej}_{j}\}$ such that when all but one ${\ej}$ are set to zero
it reduces to Eq.~(\ref{20}) and that $A_{ptb}$ defined via  $g^{ptb}$
as in Eq.~(\ref{14}) is regular on the auxiliary $CP^{1}$.\\
\underline{Regularity conditions:}\\
One can show that the regularity of $A_{ptb}$ assumes that\\
a) $g^{ptb}$ has simple poles at 
${\rho}_{\alpha}={\ae}_{\alpha}^{j}, j=1, \ldots, N$ 
where ${\ae}_{\alpha}^{j}$
are the spinors which appear in decomposition of momenta of the particles as 
in Eq.(\ref{16});\\
b) $g({\ej}_{j})^{-1}g_{ptb}$ is regular at $\rho_{\alpha}=\ae_{\alpha}^{j}$,
where $g({\ej}_{j})$ is the one-particle solution Eq.~(\ref{14}),
when only $j$-th particle ($j$-th plane wave) is present.

These conditions define  $g^{ptb}$ up to multiplication by a 
$\rho$-independent matrix on the right, that is up to the gauge freedom.
Notice that since  $g({\ej}_{j})$ depends on ${\hej}_{j}=t_{j} {\ej}_{j}$,
not on $t_{j}$ and ${\ej}_{j}$ separately, the same will be true about
$g^{ptb}$. So $g^{ptb}$ is a polynomial in  ${\hej}_{j}, j=1, \ldots, N$
with the Regularity conditions above.
To find it explicitely we use the trick called\\
\underline{Color ordering:}\\
Let us assume for a moment that the color matrixes $t_{j}, j=1, \ldots, N$
belong to a free associative algebra (no relation but 
$(t_{j}t_{k})t_{l}=t_{j}(t_{k}t_{l})$). Then $g_{ptb}$ is uniquely
represented as a sum of ordered monomials in  ${\hej}$'s:
\begin{equation}
\label{22}
g_{ptb}(\rho, \{ {\ej} \})=1+\sum_{j}g_{j}(\rho){\hej}_{j}+
\sum_{j_{1},j_{2}}g_{j_{1},j_{2}}(\rho){\hej}_{j_{1}}{\hej}_{j_{2}}+ \ldots
\end{equation}
Then the Regularity conditions  become a simple relation on the
coefficient functions $g_{j_{1},j_{2}, \ldots }$ in Eq.~(\ref{22})
which is easily solved.\\
\underline{The solution:}\\
\beq
\label{23}
g_{j_{1},j_{2}, \ldots, j_{L}}(\rho)=
\frac{(\rho,q^{j_{1}})}{(\rho,\ae^{j_{1}})}
\frac{(\ae^{j_{1}},q^{j_{2}})(\ae^{j_{2}},q^{j_{3}}) \ldots
(\ae^{j_{L-1}},q^{j_{L}})}
{(\ae^{j_{1}},\ae^{j_{2}})(\ae^{j_{2}},\ae^{j_{3}}) \ldots
(\ae^{j_{L-1}},\ae^{j_{L}})}
\eeq
Eqs.~(\ref{22}),(\ref{23}) define the solution of the problem. Of course,
it remains to be the solution after specifying the color matrixes
$t_{j}$ to obey some commutation relations.

Since $g^{ptb}$ is known, one straightforwardly finds $A_{ptb}$ via relation
Eq.(\ref{14})~\cite{RS1}. This way one describes all tree form-factors in
the self-dual sector of Yang-Mills theory. One can add one antiself-dual
plane wave solving linearization of the Yang-Mills equations in the
background of  $A_{ptb}$~\cite{RS1}.

\section{sin(h)-Gordon}
Let us now turn to the sin(h)-Gordon case (since the perturbiner is
anyway complex solution, it does not really matter whether it is
$sin$ or $sin(h)$):
\beq
\label{24}
{\partial}{\bar{\partial}}{\phi}+
\frac{m^{2}}{\beta}{\rm sinh}{\beta}{\phi}=0
\eeq
where ${\partial}=\frac{\partial}{\partial z}$, 
${\bar{\partial}}=\frac{\partial}{\partial {\bar z}}$, and $z$, ${\bar z}$
are two lightcone 
coordinates. In what follows we put $m^{2}=1$ since $m^{2}$-dependence
can easily be restored. 

According to the general definition of perturbiner, we are looking for 
solution of Eq.~(\ref{24}) of the type of Eq.~(\ref{6}), where the plane
waves are now
${\ej}_{j}=a_{j}e^{ik_{j}z+i\frac{1}{k_{j}}\bar{z}}$.

The key ingredient of the construction of perturbiner in above cases -\\
\underline{The zero-curvature representation}\\
- is very well known in the present case,
see e.g. the book~\cite{ZC}:
\begin{eqnarray}
\label{26}
A_{z}=-\frac{\beta}{4}{\sigma}_{1}{\partial}{\phi}+
\frac{\lambda}{2}{\sigma}_{3}cosh\frac{\beta{\phi}}{2}+
\frac{\lambda}{2}i{\sigma}_{2}sinh\frac{\beta{\phi}}{2}\nonumber\\
A_{\bar z}=\frac{\beta}{4}{\sigma}_{1}{\bar {\partial}}{\phi}-
\frac{1}{2 \lambda}{\sigma}_{3}cosh\frac{\beta{\phi}}{2}+
\frac{1}{2 \lambda}i{\sigma}_{2}sinh\frac{\beta{\phi}}{2}
\end{eqnarray}
where $\lambda$ is a non-homogeneous coordinate on an auxiliary $CP^{1}$ space,
the so-called spectral parameter, and ${\sigma}_{i}$ are Pauli matrixes.
The Sin(h)-Gordon equation (\ref{1}) is equivalent to
${\partial}A_{\bar z}-{\bar {\partial}}A_{z}+[A_{z},A_{\bar z}]=0$.
The connection form Eq.(\ref{26}) is meromorphic on the auxiliary $CP^{1}$
space with simple poles at $\lambda=0$ and $\lambda=\infty$. Correspondingly,
the zero-curvature condition consists in fact of a number of 
equations - at different powers of $\lambda$ - most of which are automatically
resolved when the connection form is taken in the form Eq.(\ref{26}), 
independently
of the field ${\phi}(z,{\bar z})$. The only nontrivial equation arises at
${\lambda}^{0}$ and is equivalent to Eq.(\ref{24}).\\
\underline{Mikhailov's reduction:}\\
It would be very inconvenient to look for a flat connection
of the particular form Eq.~(\ref{26}). Luckily, due to work~\cite{A.Mi}
we know how those flat connections which produce sin(h)-Gordon are
distinguished among all flat connections.
Namely, a generic zero-curvature
connection with simple poles at ${\lambda}=0, {\infty}$
obeying the ``reduction condition''
$A(-{\lambda})={\sigma}_{1}A({\lambda}){\sigma}_{1}$
is equivalent to the connection Eq.(\ref{26}) modulo gauge transformations
and a choice of coordinates $z$, ${\bar z}$. The gauge transformations are
transformations with $\lambda$-independent $SL(2,C)$ matrix commuting
with the reduction condition. 

The zero-curvature condition is (locally) solved as
$A=g^{-1}dg$
where, $g$ is a nontrivial function of $\lambda$ subject to
the condition that the connection form $A(\lambda)$ has  simple poles
at $\lambda=0$ and $\lambda=\infty$ and also that  $A(\lambda)$ obeys
the reduction condition which for $g$ gives
$g(-{\lambda})={\sigma}_{1}g({\lambda}){\sigma}_{1}$.
The gauge transformations act on $g(\lambda)$ as multiplication
by a $\lambda$-independent commuting with ${\sigma}_{1}$
matrix from the right.\\
Since ${\phi}_{ptb}$ is polynomial in the plane waves $\ej$,
so are $A_{ptb}$ and $g_{ptb}$. A novel thing compared to the Yang-Mills
case is that $A_{ptb}$ has a term of zero-th order in  $\ej$'s which is 
convenient to split off explicitely:
\begin{eqnarray}
\label{31}
 A^{ptb}(\lambda, \{{\ej}\})=A^{(0)}(\lambda)+A'(\lambda, \{{\ej}\})
\nonumber\\
A'={g'_{ptb}}^{-1}{\nabla}^{(0)}g'_{ptb},\;{\nabla}^{(0)}=d+A^{(0)}
\end{eqnarray}
where the non-derivative term in ${\nabla}^{(0)}$ acts on $g'_{ptb}$ as 
commutator.\\
Further steps of construction are parallel to the Yang-Mills case.\\
\underline{One-particle solution:}\\ 
$g'(\lambda,{\ej}_{j})=1+\frac{\beta}{4}{\ej}_{j}{\sigma}_{+}
\frac{{\lambda}+q_{j}}{{\lambda}+ik_{j}}
\frac{2ik_{j}}{ik_{j}-q_{j}}+
\frac{\beta}{4}{\ej}_{j}{\sigma}_{-}
\frac{{\lambda}-q_{j}}{{\lambda}-ik_{j}}
\frac{2ik_{j}}{ik_{j}-q_{j}}$,\\
where ${\sigma}_{\pm}=\frac{1}{2}({\sigma}_{1}{\pm}i{\sigma}_{2})$. Notice
that every particle contributes now two poles $(\lambda=\pm ik_{j})$ in $g$ 
which is intimately related with the reduction condition.

To put the problem of constructing the\\
\underline{$N$-particle solution:}\\
into a more universal form introduce now some more notations:\\
 ${\hat j}$ consisting of two indices, 
${\hat j}=(j,s); j=1, \ldots, N; s={\pm}$, notations
${\hej}_{\hat j}$, ${\hej}_{j,\pm}=\frac{\beta}{4}\frac{2ik_{j}}{ik_{j}-q_{j}}
{\ej}_{j}{\sigma}_{\pm}$, where 
${\lambda}_{\hat j}={\lambda}_{j,{\pm}}={\mp}ik_{j}$, 
$q_{\hat j}=q_{j,{\pm}}={\mp}q_{j}$ and 
$g'({\hej}_{\hat{j}})=1+
{\hej}_{\hat j}\frac{{\lambda}-q_{\hat j}}{{\lambda}-{\lambda}_{\hat j}}$.
In these notations $g'_{ptb}$ obeys just the same Regularity conditions
as $g_{ptb}$ in in Yang-Mills case so\\
\underline{The solution:}\\
for  $g'_{ptb}$ is given by Eqs.~(\ref{22}),(\ref{23}):\\
$g'(\lambda)=1+\sum_{d=1}^{N}\sum_{{\hat j_{1}}, \ldots , {\hat j_{d}}}
\frac{{\lambda}-q_{\hat j_{1}}}{{\lambda}-{\lambda}_{\hat j_{1}}}
\frac{{\lambda}_{\hat j_{1}}-q_{\hat j_{2}}}
{{\lambda}_{\hat j_{1}}-{\lambda}_{\hat j_{2}}}{\cdots}
\frac{{\lambda}_{\hat j_{d-1}}-q_{\hat j_{d}}}
{{\lambda}_{\hat j_{d-1}}-{\lambda}_{\hat j_{d}}}
{\hej}_{\hat j_{1}} {\ldots} {\hej}_{\hat j_{d}} $\\
from which one obtains, finally,\\
\beq
{\phi}^{ptb}=\sum_{d\;{\rm odd}}\frac{2}{d}(\frac{\beta}{2})^{d-1}
\sum_{{j_{1}}, \ldots , {j_{d}}}
\frac{k_{j_{1}} {\ldots} k_{j_{d}}}
{(k_{j_{1}}+k_{j_{2}}) {\ldots} (k_{j_{d}}+k_{j_{1}})}
{\ej}_{j_{1}} {\ldots} {\ej}_{j_{d}}.
\eeq

\section{Gravity} 
Due to lack of space and time I am not able to say anything about 
gravitational perturbiner, I just refer to the original 
works~\cite{RS2}$^{-}$~\cite{KS4}.

\section*{Acknowledgments}
I would like to acknowledge RFBR grant 99-01-10584 and the organizers of 
the conference for financial support. I would also like to thank the 
organizers for their kind hospitality and for the beautiful
surroundings of the conference.

\end{document}